\documentclass{PoS}

\usepackage{graphicx}
\usepackage{amsmath}

\let\Re\relax
\DeclareMathOperator{\Re}{Re}
\let\Im\relax
\DeclareMathOperator{\Im}{Im}
\DeclareMathOperator{\Tr}{Tr}

\renewcommand{\d}{\mathrm{d}}

\title{Beyond Thimbles: Sign-Optimized Manifolds for Finite Density}

\ShortTitle{Sign-Optimized Manifolds for Finite Density}

\author{\speaker{Scott Lawrence}\\
        Department of Physics, University of Maryland, College Park, MD 20742, USA\\
        E-mail: \email{srl@umd.edu}}

\abstract{The sign problem of relativistic field theories at finite fermion chemical potential has been approached by deforming the domain of integration into complex field space. We present a method for selecting a deformed manifold of integration which is a local maximum of the average phase, and demonstrate this method on the three-dimensional Thirring model. Finally, we compare the performance of this method, in the heavy-dense limit, to direct integration on the Lefschetz thimbles.}

\FullConference{The 36th Annual International Symposium on Lattice Field Theory - LATTICE2018\\
		22-28 July, 2018\\
		Michigan State University, East Lansing, Michigan, USA.}

\begin{document}

\section{Introduction}
Lattice Monte Carlo methods, the only broadly applicable approach to studying strongly coupled field theories, compute thermal expectation values via importance sampling according to the Boltzmann factor $e^{-S(\phi)}$, where $S$ is the euclidean-spacetime lattice action and $\phi$ is a field configuration. In fermionic theories with a non-zero chemical potential, the action is typically not real, so that $e^{-S}$ cannot be interpreted as a probability distribution. This is termed the \emph{fermion sign problem}. The severity of the sign problem is characterized by the average phase
\begin{equation}\label{eq:average-phase}
	\left<\sigma\right> \equiv \frac{\int \mathcal D \phi\;e^{-S(\phi)}}{\int \mathcal D \phi\;e^{-\Re S(\phi)}}\text,
\end{equation}
where $\left<\sigma\right>=1$ in the absence of a sign problem, and the average phase becomes smaller as the sign problem becomes more severe. Typically, the average phase decays exponentially with the spacetime volume of the system being simulated. In the presence of a sign problem, expectation values are evaluated by reweighting:
\begin{equation}
	\left<\mathcal O\right> =
	\frac
	{\left.\int\mathcal D \phi\;\mathcal O(\phi) e^{-S(\phi)}\middle/\int\mathcal D \phi\;e^{-\Re S(\phi)}\right.}
	{\left.\int\mathcal D \phi\;e^{-S(\phi)}\middle/\int\mathcal D \phi\;e^{-\Re S(\phi)}\right.}
	=
	\frac
	{\left<\mathcal O e^{-i \Im S}\right>_{\Re S}}
	{\left<e^{-i \Im S}\right>_{\Re S}}
	\text.
\end{equation}
Here $\left<\mathcal O\right>_{\Re S}$ denotes an expectation value taken with respect to the probability density $e^{-\Re S}$. The denominator is the average phase $\left<\sigma\right>$. The smallness of the denominator increases the noise in the estimator, necessitating more samples to achieve the same statistical error.

Towards alleviating the sign problem, it was proposed to view the path integral as a contour integral (in the sense of complex analysis), and deform the integration contour into complex field space. Initially, the deformed integration contour was chosen to be the Lefschetz thimbles \cite{Cristoforetti:2012su}. Later work proposed manifolds interpolating between the real plane and the thimbles \cite{Alexandru:2015sua}, and used machine learning methods to obtain a more efficient parameterization \cite{Alexandru:2017czx}. While these approaches have met with some success, an efficient algorithm for integrating on the thimbles remains elusive. Additionally, it is generally the case that the Lefschetz thimbles are not the manifold with the mildest-possible sign problem --- we expect a more judicious choice of manifolds to do better.

In this talk, we present a method \cite{Alexandru:2018fqp} for optimizing $\left<\sigma\right>$ in a large class of manifolds. This method is used to compute the phase diagram of a lattice Thirring model in $2+1$ dimensions \cite{Alexandru:2018ddf}. Similar methods have been proposed and applied to two-dimensional scalar field theory \cite{Mori:2017nwj}, QCD effective theories \cite{Kashiwa:2018vxr}, and a one-dimensional bose gas \cite{Bursa:2018ykf}.

As a testbed for this method, we will consider the Thirring model in $2+1$ dimensions, defined by the euclidean lattice action
\begin{equation}\label{eq:lattice-action}
S=\sum_{x,\nu} \frac{N_F}{g^2} (1-\cos A_\nu(x))+\sum_{x,y} \bar\psi^a(x) D_{xy}(A)  \psi^a(y)
\end{equation}
where $-\pi<A_\mu(x)\le\pi$ is a compact bosonic auxiliary field.  The staggered fermion matrix is given by
\begin{equation}
D_{xy} = m\delta_{xy} + \frac{1}{2}\sum_{\nu=0}^2  
\Big[ 
 \eta_{\nu}(x) e^{i A_\nu(x)+\mu \delta_{\nu 0}} \delta_{x+\hat\nu, y}
 -\eta^\dag_{\nu}(y)e^{-i A_\nu(y)-\mu \delta_{\nu 0}}  \delta_{x, y+\hat\nu}
\Big]\text,
\end{equation} 
where $\eta_\nu(x)=(-1)^{x_0+\ldots+x_{\nu-1}}$,
%$\eta_0=1$, $\eta_1=(-1)^{x_0}$, $\eta_2=(-1)^{x_0+x_1}$ 
the flavor index $a$ takes values from $1,\ldots,N_F/2$, $g$ is the coupling, $\mu$ is the chemical potential, and $m$ is the bare mass. We work with $N_F=2$ flavors of fermions.

\section{Path Integral as a Contour Integral}

Before deforming the domain of integration of the path integral, we must define the complexified field space. For a three-dimensional lattice with $V$ sites, there are $3V$ compact degrees of freedom $A_\mu(x)$, so that the field space is the hypertorus $\left(S^1\right)^{3V} \approx \mathbb T^{3V}$. The lattice path integral is originally an integral over this torus. The complexification of $S^1$ is the complex plane without the origin $\mathbb C \backslash \{0\}$, or equivalently, the cylinder $S^1 \times \mathbb R$. The complexified $A_\mu(x)$ live in the latter space; the full complexified field space is $\left(S^1 \times \mathbb R\right)^{3V}$. The deformed integration contour will be a $3V$-dimensional hypersurface in this space.

The deformation of the contour of integration is enabled by Cauchy's integral theorem, which states that if a function $f$ is holomorphic on an $N+1$-dimensional region $\Omega \subset \mathbb C^N$, then the integral along the boundary vanishes: $\int_{\partial\Omega} f\;\d z = 0$. As a direct consequence, if an $N$-real-dimensional hypersurface $\gamma_1$ can be deformed continuously into another $\gamma_2$, the integrals on the two surfaces are equal. In particular, if a manifold $\mathcal M$ can be obtained by a continuous deformation of the original domain of integration $\mathbb T^{3V}$, then
\begin{equation}
	\int_{\mathcal M} f(\tilde\phi) \; \d^{3V}\!\tilde\phi
	=
	\int_{\mathbb T^{3V}} f(\phi) \; \d^{3V}\!\phi
\end{equation}
for any holomorphic function $f(\phi)$.

In practice, the integration along $\mathcal M$ is performed by parameterizing $\mathcal M$ by the original, real domain. A continuous function $\tilde\phi(\phi)$ is constructed which takes a point $\phi \in \mathbb T^{3V}$ to a point $\tilde\phi(\phi) \in \mathcal M$. An integral along $\mathcal M$ can now be written
\begin{equation}\label{eq:integrate-curved}
	\int_{\mathcal \phi} \mathcal D \tilde\phi\;e^{-S(\tilde\phi)}
	=
	\int_{\mathbb T^{3V}} \mathcal D \phi\;e^{-S[\tilde\phi(\phi)] + \log\det J}
	\text,
\end{equation}
where $J_{ij} \equiv \frac{\partial \tilde\phi_i}{\partial \phi_j}$ is the jacobian introduced by the change of variables.

For physical observables to be unchanged, it is necessary that all integrands of physical interest are holomorphic functions. In the case of the Boltzmann factor, this is easily seen, as the fermionic and bosonic pieces may be separated as $e^{-S_B} \det D$ into a product of two manifestly holomorphic factors. It is less clear that integrands involved in propagators, for instance, are holomorphic: in the meson propagator
\begin{equation*}
	\left<\bar\psi_i\bar\psi_j\psi_j\psi_i\right> = \frac 1 Z
	\int \mathcal D \phi\; e^{-S_B(\phi)} \det D \left[ D^{-1}_{ij} D^{-1}_{ji} - D^{-1}_{ii} D^{-1}_{jj}\right]\text,
\end{equation*}
the integrand appears to be singular where $\det D = 0$. However, as shown in the appendix of \cite{Alexandru:2018ngw}, all such integrands arising from fermionic many-point functions are indeed holomorphic.

Note that although the integrands involved in estimating physical observables are holomorphic, the integrand in the numerator of (\ref{eq:average-phase}) is not. This implies that the average phase will depend on the manifold chosen for integration, and justifies this approach to resolving the sign problem.

\section{Optimizing the Manifold}
Given a family of manifolds $\mathcal M_\lambda$, we will use gradient ascent to find a local maximum of the average phase $\left<\sigma\right>$. To do so efficiently, we exploit the fact that the numerator of (\ref{eq:average-phase}), being the integral of a holomorphic function, does not depend on $\lambda$. The gradient of the average phase, then, is determined entirely by $\nabla_\lambda \int e^{-\Re S}$.  As a derivative of a \emph{phase-quenched} partition function, this may be written as a phase-quenched observable --- in particular, it has no sign problem. Working with the real part of $\log\left<\sigma\right>$, we find
\begin{equation}\label{eq:gradient}
	\nabla_\lambda \log \left|\left<\sigma\right>_\lambda\right|
	=
	\left<\nabla_\lambda \Re S - \Re \Tr J^{-1} \nabla_\lambda J\right>_{\Re S}\text.
\end{equation}

The gradient ascent begins at an arbitrary point $\mathcal M_{\lambda_0}$ in manifold space. At each step, a short Monte Carlo is done to estimate $\nabla_\lambda \log \left|\left<\sigma\right>_\lambda\right|$, and then $\lambda$ is updated according to
\begin{equation}
	\lambda_{i+1} = \lambda_{i} + \epsilon \nabla_\lambda \log \left|\left<\sigma\right>_\lambda\right|
	\text.
\end{equation}
This process is repeated until the parameters $\lambda$ converge, at which point an ordinary Monte Carlo is performed on the manifold $\mathcal M_{\lambda_{\text{final}}}$ to determine physical expectation values.

This algorithm is enabled by the fact that expression (\ref{eq:gradient}) for the gradient of the average phase can be computed efficiently. This is a somewhat remarkable fact: the average phase itself is difficult to estimate, but the direction in which a manifold should be deformed to improve the average phase is relatively easy to estimate.

In practice, the gradient ascent process is made more efficient by the use of adaptive stepsize, momentum, and other improvements. For the results in this talk, \textsc{Adam} \cite{2014arXiv1412.6980K} was used.

\section{Heavy-Dense Limit}\label{sec:heavy-dense}

The procedure described thus far is useful only if we have an ansatz for a manifold on which the average phase $\left<\sigma\right>$ may be tolerably large. In the case of the Thirring model, we produce such an ansatz by considering the heavy-dense limit. We expand the fermion determinant in powers of $e^{-\mu}$, and consider the leading term. At large $\mu$, where the average phase is smallest, this is the term that dominates. In this limit, the fermion determinant is simply $\det D = e^{\beta L^2 \mu}\left(e^{i \sum_x A_0(x)} + \mathcal O(e^{-\beta\mu})\right)$. Therefore, the lattice path integral reads
\begin{equation}\label{eq:heavy-dense}
	Z = \left[\int \d A_0 \;e^{\frac {N_F} {g^2} \cos A_0 + \mu + i A_0}\right]^{\beta L^2} \left[\int \d A_1\; e^{\frac {N_F} {g^2} \cos A_1}\right]^{2 \beta L^2}\text.
\end{equation}
In this way the path integral factorizes into a product of $3 \beta L^2$ independent integrals, one for each lattice degree of freedom. Motivated by this observation, we chose as an ansatz a manifold which also factorizes: the imaginary part of $A_\mu(x)$ is a function only of the real part of $A_\mu(x)$, depending on no other degrees of freedom. The symmetry $A \rightarrow -A$ of the action suggests that we restrict ourselves to manifolds for which the imaginary part is an even function of the real part. Finally, as the spatial integral (over $A_1$) has no sign problem, we deform only $A_0$. A general manifold in this class is given by the fourier series
\begin{align}\label{eq:ansatz}
	\tilde A_0(x) &= A_0(x) + i \left[\lambda_0 + \lambda_1 \cos A_0(x) + \lambda_2 \cos 2 A_0(x) + \cdots\right]\text,\\
	\tilde A_i(x) &= A_i(x)\text.\label{eq:ansatz2}
\end{align}
We truncate the series after the third term, so that the family of manifolds has only the three parameters $\{\lambda_0, \lambda_1, \lambda_2\}$.

Because this manifold ansatz factorizes, the jacobian matrix in (\ref{eq:integrate-curved}) is diagonal. This is computationally convenient, as computing $\det J$ for a generic matrix $J$ is an $O(V^3)$ task, where $V$ is the spacetime volume of the lattice. For a diagonal matrix, this requires only $O(V)$ operations.

\section{Lattice Calculation}

The ansatz given by (\ref{eq:ansatz}) and (\ref{eq:ansatz2}), after optimization, allows us to access substantially larger chemical potentials than would have been feasible on the real plane. We simulate $\beta \times 6^2$ lattices for $\beta =4,6,8,10,12$, using a bare mass of $m = 0.01$ and a coupling constant $g = 1.08$. These parameters are chosen to bring the lattice near a chiral phase transition. We quote the results of these calculations in lattice units.

On a spatial volume of $10^2$, we measure the fermion and meson masses to be $m_f = 0.46(1)$ and $m_b = 0.21(1)$, respectively. These masses have a small dependence on the volume $L^2$, but at all volumes we find $m_b \ll 2 m_f$, indicating a strongly coupled theory.

\begin{figure}
	\centering
	\includegraphics[width=0.48\textwidth]{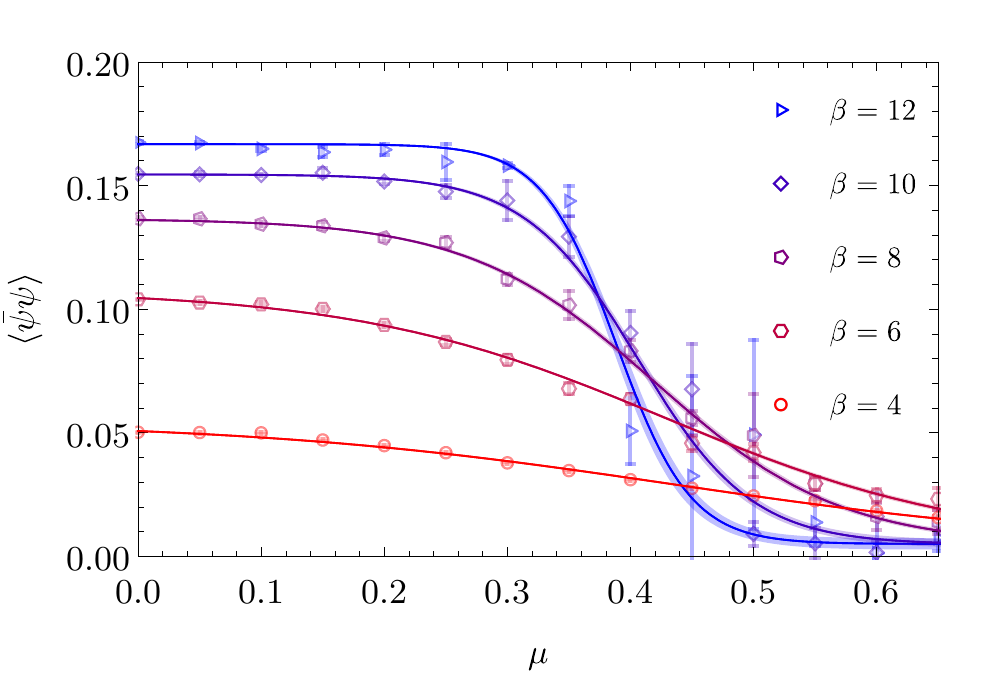}
	\includegraphics[width=0.48\textwidth]{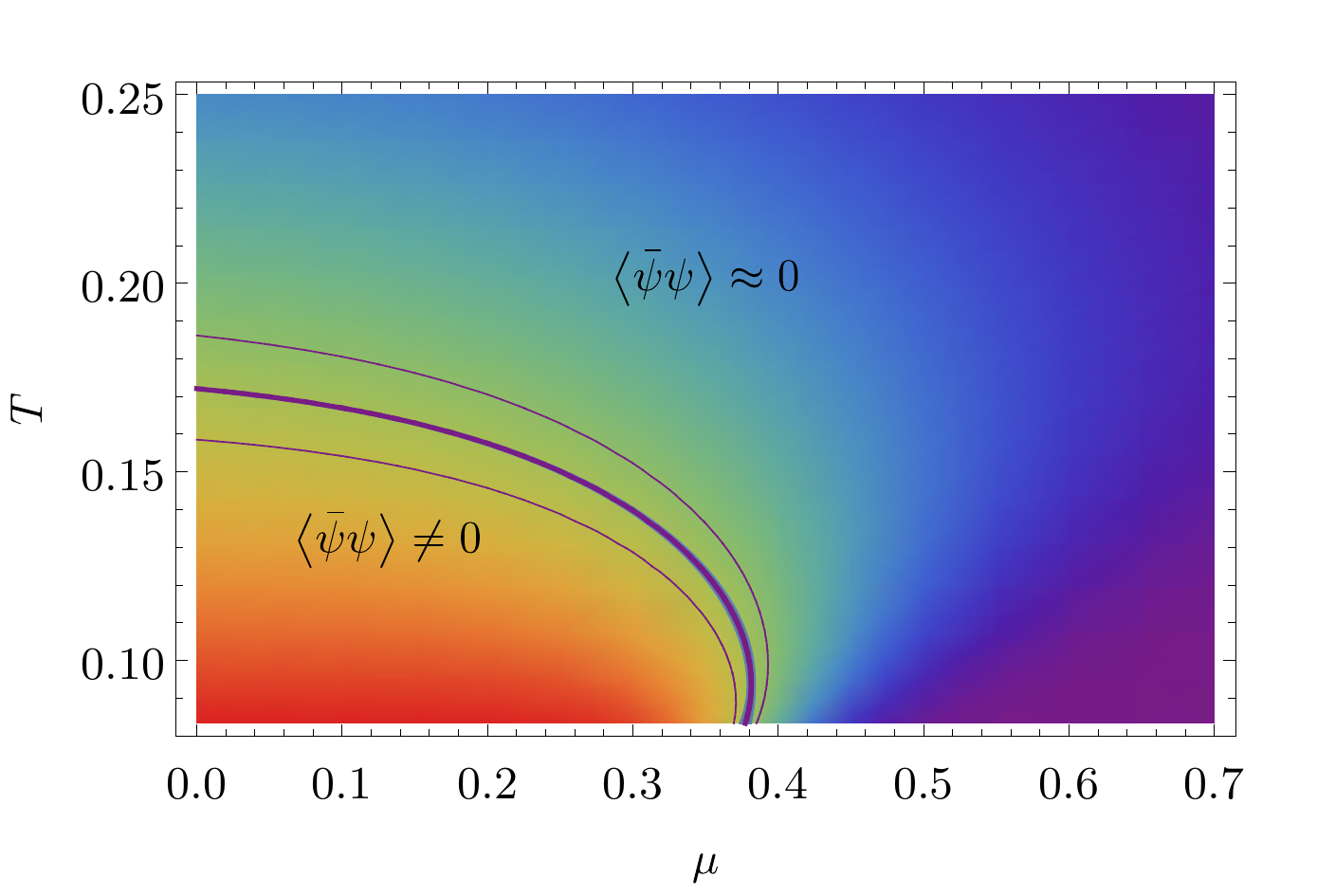}
	\caption{The chiral symmetry breaking phase transition. On the left, the condensate $\left<\bar\psi\psi\right>$, as a function of chemical potential $\mu$ on $\beta\times 6^2$ lattices, showing the condensate melting (and symmetry being approximately restored) at large densities. On the right, the full $T$-$\mu$ plane for the same spatial volume. The central band indicates the location of $\left<\bar\psi\psi\right>_{\mu,T} = 0.5 \left<\bar\psi\psi\right>_{0}$. The side bands mark $\left<\bar\psi\psi\right>_{\mu,T} = \left(0.5 \pm 0.05\right)\left<\bar\psi\psi\right>_0$, indicating the sharpness of the transition. Details of the fits done to determine these contours are given in \cite{Alexandru:2018ddf}.\label{fig:transition}}
\end{figure}

The chiral condensate $\left<\bar\psi\psi\right>$ is computed across a range of chemical potentials, and plotted in Fig.~\ref{fig:transition}, along with the resulting phase diagram. The condensate can be seen to melt, marking the restoration of approximate chiral symmetry, at both large temperatures and large chemical potential.

\section{Heavy-Dense Lefschetz Thimbles}

Despite great advances of technology over the last decade, the task of integrating directly on the Lefschetz thimbles remains well out of reach in the general case. We would nevertheless like to compare the average phase on the sign-optimized manifolds used here to that on the thimbles. This will compare the algorithm described in this talk to a hypothetical, efficient algorithm for integrating exactly on the Lefschetz thimbles. No such general algorithm exists; however, in certain limits where the thimbles are exactly known, this comparison is possible.

One such limit is $\mu = 0$; however, this limit is trivial, as neither the manifold ansatz described here, nor the thimbles have a sign problem at all. More significant is the heavy-dense limit, in which the partition function factorizes into $\beta V$ identical one-dimensional integrals: $Z = Z_1^{\beta V}$.

Because the partition function factorizes, the average phase does as well. The average phase on a spacetime volume $\beta V$ is exactly the average phase on a single site, raised to that power: $\left<\sigma\right> = \left<\sigma\right>_1^{\beta V}$. This relation holds both on the ansatz and on the thimbles themselves. Therefore, a comparison of the ansatz and the thimbles requires taking only one-dimensional integrals to compute $\left<\sigma\right>_1$, from which $\left<\sigma\right>$ at any volume and temperature can be easily obtained.

Performing this comparison with a coupling of $g = 1.08$, we find that the average phase on the real plane is $0.645$, the average phase on the thimbles is $0.985$, and the average phase on the sign-optimized manifold is $0.9996$. These differences are small on a single-site lattice: a calculation on the thimbles would take only $3\%$ longer than one on the sign-optimized ansatz. However, the difference is overwhelming on a more substantial lattice. On the largest lattice simulated ($10^3$ sites), the real-plane average phase is $4 \times 10^{-191}$, the thimble average phase is $3 \times 10^{-7}$, and the ansatz average phase is $0.67$. The ansatz has a speed advantage of a factor of $5 \times 10^{12}$ over the thimbles, before the efficiency of integration itself is taken into account.

\begin{figure}
	\centering
	\includegraphics[height=2.4in]{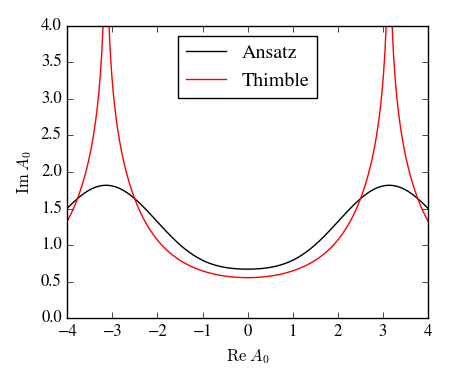}
	\caption{The thimble and ansatz manifold compared in the $A_0$ direction. The curvature of the thimble (red) results in large cancellations not present on the ansatz manifold (black).\label{fig:thimbles}}
\end{figure}

To understand the origin of the thimbles' disadvantage, the thimbles and ansatz manifold are compared in Fig.~\ref{fig:thimbles}. Near the minimum of the action, at $\Re A_0 = 0$, the two manifolds are quite similar. However, near $\Re A_0 = \pi$ the thimbles curve sharply upward, introducing large phase cancellations. The sign-optimized manifold lacks these cancellations, resulting in a larger average phase.

This comparison establishes that, even on a relatively tame $10^3$ lattice, calculations on the Lefschetz thimbles are unable to access certain limits of the theory, while other integration contours are able to do so.

\section{Conclusions}

The method of sign-optimized manifolds described here allows a large space of trial manifolds to be efficiently explored by gradient ascent, in a search for the manifold with the largest average phase, and therefore the mildest sign problem. A particular ansatz was constructed for the Thirring model, and demonstrated in $2+1$ dimensions. A large section of the $T$-$\mu$ plane may be explored with the resulting manifolds, demonstrating the melting of a chiral condensate at large fermion densities.

In the limit of large chemical potential $\mu$, this method can be compared directly to an integration directly on the Lefschetz thimbles; the sign-optimized manifolds exhibit an average phase many orders of magnitude larger than that on the thimbles. This is a concrete demonstration of the fact that Lefschetz thimbles do not provide the best-possible average phase, and for high-dimensional problems, better manifolds are required.

Future work must focus on exploring larger classes of manifolds \cite{Mori:2017nwj,Bursa:2018ykf}, with an eye towards finding manifolds with tolerable average phases on gauge theories and theories in $3+1$ dimensions. 

\begin{acknowledgments}
This work was done in collaboration with Andrei Alexandru, Paulo Bedaque, Henry Lamm, and Neill Warrington.
S.L. is supported by the U.S. Department of Energy under Contract No.~DE-FG02-93ER-40762.
\end{acknowledgments}

\bibliographystyle{JHEP}
\bibliography{thimbology}

\end{document}